# Quantifying quantum risk: a measure of crypto agility

*Coryan Wilson-Shah*

## Abstract

Because of their ability to enable new forms of cryptanalysis, quantum computers pose a threat to the cryptographic algorithms that are widely used to secure contemporary computer systems. A practical quantum computer may emerge within the next ten years or so, but due to theorised "harvest now, decrypt later" style attacker behaviour, mitigations are necessary today. Recent advances in cryptography and security architecture show promise in supporting the design of systems that exhibit resilience against quantum-enabled cryptanalysis, however there is a key gap in the literature around the subject of deriving tolerances for such systems. In this paper, we introduce the concept of rotation time as a measure of crypto agility, and derive an approximation that links rotation time tolerance to security risk tolerance. Historical CVE data is used to calculate illustrative values for rotation time tolerance, which is found to be of the order of hours to days. This demonstrates that using crypto agility in conjunction with hybrid encryption is an effective approach for designing quantum-resilient systems, but may necessitate challenging technical and operational tolerances in order to meet organisational risk tolerances.

---

## Context

Recent years have seen accelerating developments in the field of quantum computing, and a practical, commercially available cryptographically relevant quantum computer (CRQC) [1] now seems closer to becoming a reality [2] [3] [4] [5]. This has prompted concerns regarding the resilience of widespread information security controls that rely upon contemporary cryptography, since a CRQC might be used to reduce their effectiveness [6]. RSA, for example, derives its security properties from the difficulty associated with factorising the products of large prime integers [7], and AES uses a construction that provides resistance against brute-force attacks [8]. Quantum algorithms enable an attacker to undermine the security of both: Shor's algorithm enables efficient integer factorisation [9], thereby compromising RSA private keys, whereas Grover's algorithm makes for more efficient brute-forcing of AES-encrypted data [10], reducing AES's effective key length. A lack of access to a practical CRQC has meant that these attacks have remained purely theoretical, however the current trajectory of quantum hardware development indicates that they may become practically feasible within the next decade or so [11] [12]. It has been theorised that, even in the absence of a practical CRQC, today's attackers may be using a "harvest now, decrypt later" approach – also known as retrospective decryption – to collect encrypted data and store it until such a time as a quantum-enabled cryptanalysis capability is available to them [13].

The threat of quantum-enabled cryptanalysis has necessitated urgent exploration of post-quantum cryptography and mitigation strategies. A set of post-quantum algorithms has been standardised [14] [15], though these remain largely untested at scale and their long-term reliability is not yet established [16]. This leaves defenders with a difficult choice: adopt emerging post-quantum algorithms, thereby accepting the risk associated with relying on algorithms that are untested at scale, or continue using contemporary cryptography, thereby accepting the risk that adversaries will at some point have access to a quantum-enabled cryptanalysis capability. For many organisations, neither option will appear palatable. There is a high cost associated with getting this choice wrong: migrating to post-quantum algorithms can be a long process spanning several years. [1] [17] [18]

In response to this challenge, two complimentary strategies have been suggested to make cryptographic systems more resilient to attack by quantum-enabled cryptanalysis. The first of these, **hybrid encryption**, involves the use of two distinct algorithms in tandem to mitigate the risk associated with a single point of failure [19] [20] [21]. The second, **crypto agility**, involves designing systems with the ability to rapidly reconfigure their cryptographic controls [22] [23] [24]. Hybrid encryption effectively treats the problem of choosing between contemporary or post-quantum algorithms by instead using both in tandem (sometimes referred to as **double encryption**.) This grants the benefits of both approaches: the quantum resistance of post-quantum algorithms, along with the reliability of contemporary algorithms. Crypto agility reduces the cost associated with migration between different cryptographic controls: it acknowledges that, sooner or later, vulnerabilities will be discovered in any security control, be they algorithmic weaknesses, newly discovered implementation flaws, or the actualisation of a quantum-enabled cryptanalysis capability. It calls for the use of architectures that have the ability to reconfigure their cryptographic controls with minimal friction, allowing systems to rapidly reconfigure in response to emergent vulnerabilities - quantum-enabled or otherwise. The use of these two strategies is supported by recent work [25] [26], which emphasise the operational challenges and increased complexity associated with effective quantum-resistant security architecture.

Hybrid encryption and crypto agility are both architectural choices: they are design-level controls that enhance the resilience of a system against emergent vulnerabilities. They do not, however, necessarily provide all the details necessary to define a set of technical requirements or an operational approach that will remain resilient against emergent vulnerabilities. They raise a number of important questions:

- How agile is agile enough?
- How do we measure agility?
- How agile must a system be in order to be considered crypto agile?

There is no clear view on how to align the specification of a crypto agile, hybrid encrypted system with a given risk tolerance, and the literature highlights a persistent gap between risk models and real-world crypto agile/hybrid encryption practices [24].

This paper seeks to address this gap by providing a quantitative model that links risk tolerance to technical specifications for a crypto agile, hybrid encrypted system, thereby enabling informed decisions around design tolerances. We introduce the concept of rotation time and use it to quantify the crypto agility of a system. We then show how to approximate acceptable upper bounds on rotation time for a system with a stated risk tolerance.

## System modelling

To assist in our reasoning, we define the **rotation time** as the time taken for a crypto agile system to patch, replace, or otherwise swap out a cryptographic control. Rotation time provides a means by which to reason about a system's agility and provides an answer the question "how do we measure agility" – rotation time can be thought of as a measure of the crypto agility of a system.

The question of "how agile is agile enough" can then be phrased as:

**For a system with a given risk tolerance, what is the maximum tolerable rotation time?**

To answer this question, we will need a simple model of a representative system. Let's model a simple system that uses the mitigation strategy outlined earlier: crypto agility in conjunction with hybrid encryption. This system has the following characteristics:

- The system has a single asset, which must be secured from an attacker.
- The system uses two different cryptographic controls, $C_1$ and $C_2$, to secure the asset from unauthorised access.
    - These two controls use different algorithms, different implementations, and different key material.
    - Each control uses a proven, peer-reviewed algorithm that has no known algorithmic weaknesses and is widely adopted and standardised, and a proven, peer-reviewed implementation that has no known weaknesses and is widely adopted and standardised.

- In order to compromise the asset, the attacker must simultaneously breach both $C_1$ and $C_2$.
- The system is under constant attack by a motivated adversary; as soon as the attacker is able to exploit a vulnerability in our system, they will do so.
    - This might be less realistic for attackers using n-days, as they might need to build and test a weaponised exploit - but for attackers with access to 0-days, this starts to look a bit more representative. It is prudent for us to assume the worst-case scenario, which is a highly capable attacker.
- The attacker and defenders possess equivalent knowledge of vulnerabilities; once a vulnerability is publicly disclosed, both parties are immediately aware of it.

Since we are using robust, peer-reviewed algorithms with robust, peer-reviewed implementations, we can generally assume that our cryptographic controls are secure until a new weakness is discovered.

From practice, it is reasonable to assume the following statements regarding vulnerabilities in established cryptographic controls:

- They occur at approximately a constant average rate.
- They are discrete and count based: there is either a vulnerability, or there isn't.
- The probability of more than one different vulnerability being disclosed in a small interval is small (this isn't quite true, as researchers often publish several related CVEs together, but it's not an absurd approximation.)
- Each vulnerability is independent of others (this is also not quite true in practice, but also not an absurd approximation to make for the sake of a convenient model.)

It is important to recognise that these assumptions are simplifications: other, more comprehensive vulnerability discovery models have been proposed that do not rely on these assumptions [27] [28] [29]. They nonetheless remain practical and effective simplifications that will aid us in modelling our system.

Along with these simplifications, we also hold that, from a vulnerability management perspective, the emergence of a practical CRQC – and therefore the emergence of practical quantum-enabled cryptanalysis – is no different from any other emergent exploitable vulnerability. Both lead to a similar outcome (a degradation of system security) and warrant the same response (rapid mitigation), even if the underlying vulnerability has no prior equivalent. This observation is useful for informing our approach to mitigating quantum-enabled vulnerabilities, as it enables us to reason about their impact using data about past emergent (but not quantum-enabled) vulnerabilities.

Given the simplifications above, we can use a simplified logarithmic Poisson model [30] to model the vulnerabilities *U* and *V* (quantum enabled or otherwise) that occur in either one of our two cryptographic controls $C_1$ and $C_2$ over a fixed time period as Poisson variables, i.e.

$$U \quad Poisson(\lambda_U)$$

$$V \quad Poisson(\lambda_V)$$

Where $\lambda_u$ and $\lambda_v$ are the Poisson rates per unit time for each of *U* and *V* [31]. Given that $C_1$ and $C_2$ use different algorithms, different implementations, and different key material, we also make the reasonable assume that *U* and *V* are independent. The values $u_1$, $u_2$, $u_3$… are the point times at which a vulnerability is disclosed in $C_1$, and $v_1$, $v_2$, $v_3$… are the point times at which a vulnerability is disclosed in $C_2$.

Our system also has an agreed risk tolerance *R*. In practice, risk tolerances are expressed in a number of ways [32], including the following:

- Maximum acceptable likelihood of compromise
- Maximum expected loss (e.g. cost impact of loss × maximum acceptable likelihood of compromise)

The second of these is defined in terms of the first, so to keep our model as general as possible, we define R to be expressed as the maximum acceptable likelihood that the system's asset is compromised in a year. When designing security for our system, we need to apply controls such that the real likelihood of our asset being compromised is equal to or lower than the risk tolerance R:

$$P(compromise) \leq R, R \in ¿$$

We therefore need to find an expression for *P(compromise)*.

## Scenario modelling

Let's have a closer look at how this system might fail. Earlier we established that:

- In order to compromise the asset, the attacker must simultaneously breach both $C_1$ and $C_2$.

Let us consider a scenario in which a vulnerability has been disclosed that affects one of our cryptographic controls, $C_1$. We assume that our motivated adversary is immediately able to breach the vulnerable control $C_1$, and is now seeking to exploit $C_2$, thereby compromising the system's single asset.

One of two outcomes might ensue: a "secure" outcome, or a "compromise" outcome

- Secure: After some time $t_{sec}$ – which is the rotation time - we are able to replace or patch $C_1$, securing the asset from our attacker.
- Compromise: After some time $t_{comp}$, the attacker is able to exploit $C_2$, thereby fully compromising the asset.

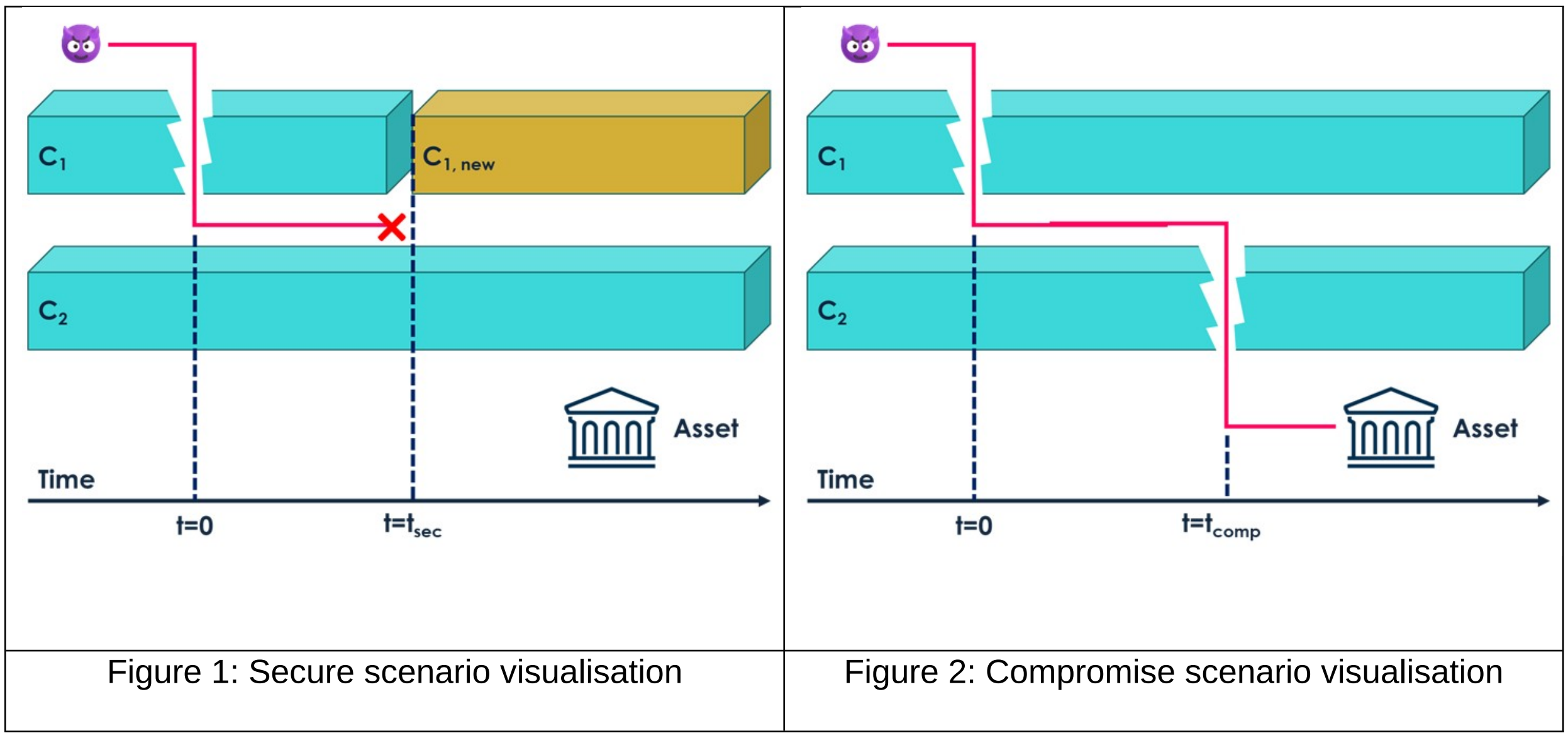

| Figure 1: Secure scenario visualisation | Figure 2: Compromise scenario visualisation |
|---|---|

It is clear that in order for the compromise outcome to occur, the attacker will need to exploit $C_2$ before we are able to patch or replace $C_1$. Since the time it takes for us to patch or replace $C_1$ is $t_{sec}$, the attacker will need to find at least one vulnerability in $C_2$ within the period $t_{sec}$, i.e. $t_{comp} < t_{sec}$. It should also be clear that, since $C_1$ and $C_2$ are independent, we can follow a similar argument with $C_1$ and $C_2$ swapped: if the attacker were to first breach $C_2$, they would then need to then breach $C_1$ within $t_{sec}$ in order to compromise the system's asset.

## Approximating probability of compromise

It follows that the probability of the compromise outcome is simply the probability that, over a time period *T* (which in this case, we're taking to be one year) there exists at least one pair (*u, v*) for which the intervening time between *u* and *v* is less than or equal to some time *t*.

Let's start by defining S as the number of times that such a pair (*u*, *v*) occurs within the time period *T*. The exact distribution of *S* is complicated by the possibility of overlapping pairs. However, if pairs are rare, we can approximate S as a Poisson variable, i.e.:

$$S \quad Poisson(\lambda_S)$$

The probability of compromise is the probability that *S* is greater than or equal to 1, i.e.:

$$P(compromise) = P(S \geq 1)$$

$$¿ 1 - P(S=0)$$

$$¿1-e^{-\lambda_s}$$

We can calculate this for *S* if we know $\lambda_S$, the expected value of *S* per unit time. Our next step must therefore be to find an expression for $\lambda_S$.

Over a small time period $\Delta u$, the expected number of occurrences of *U* is $\lambda_U \Delta u$, and the number of occurrences of *V* is $\lambda_V \Delta v$. Over a small time period, the expected value of *S* is therefore:

$$\Delta E(S) = \lambda_U \lambda_V \Delta u \Delta v$$

We can calculate $E(S)$ over our time period *T* by summing these values over *T*, including only values for which $|u-v| \leq t$, i.e.:

$$E(S) = \int_{u=0}^{T} \int_{v=0}^{T} \lambda_U \lambda_V \, du dv, |u-v| \leq t$$

A geometric solution is provided to aid intuition.

We plot *u* against *v*, and shade the area in which $|u-v| \leq t$:

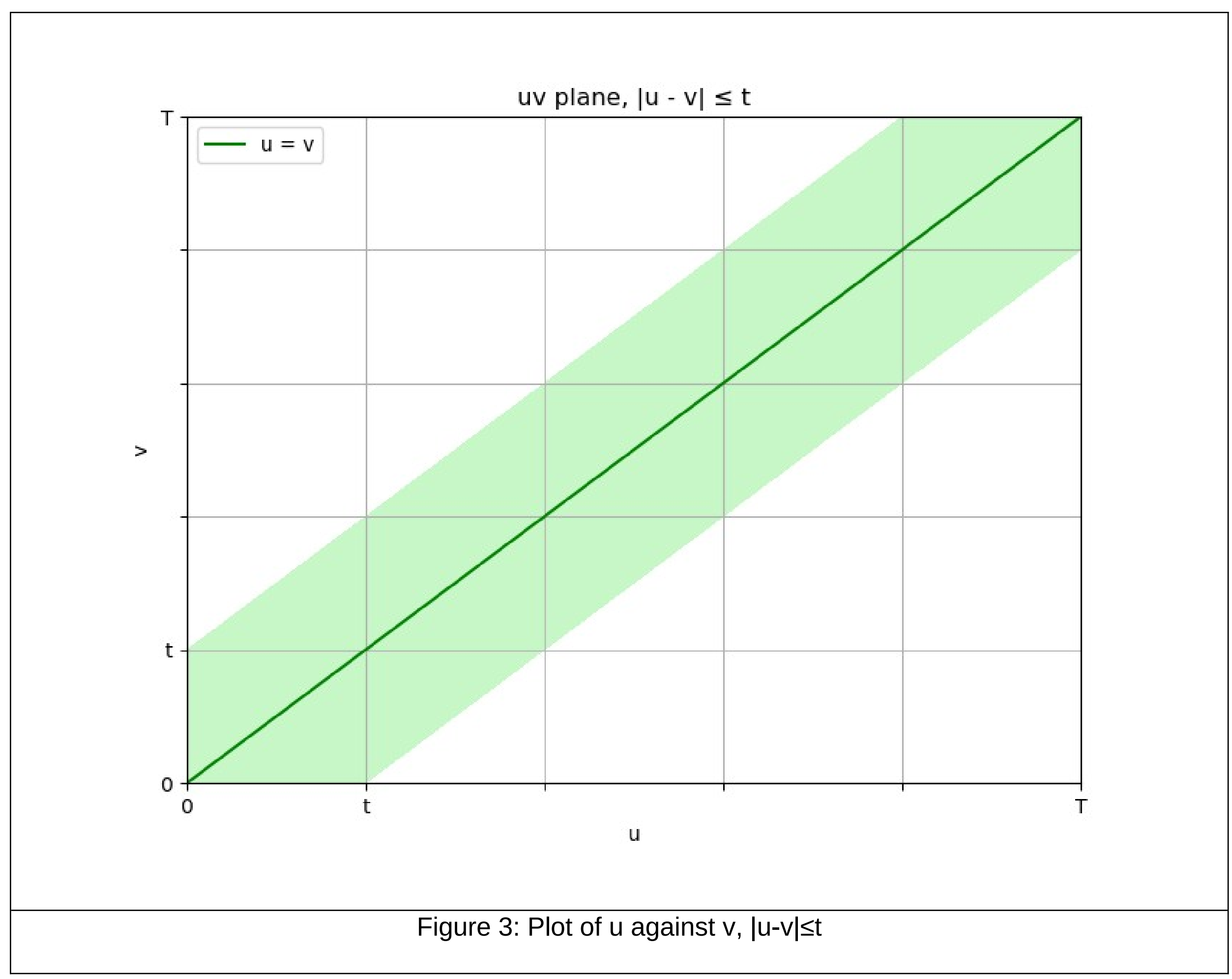

Figure 3: Plot of u against v, |u-v|≤t

We can calculate *E(S)* as $\lambda_U \lambda_V$ multiplied by the area of this shaded shape. That area is $2Tt - t^2$:

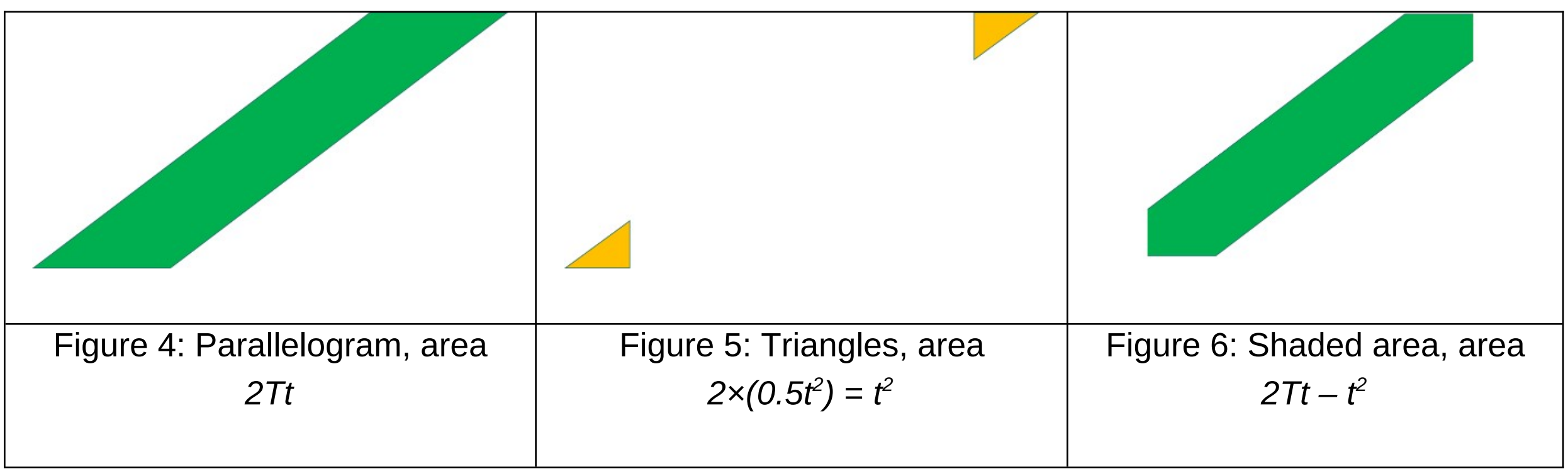

| Figure 4: Parallelogram, area *2Tt* | Figure 5: Triangles, area *2×(0.5t²) = t²* | Figure 6: Shaded area, area *2Tt – t²* |
|---|---|---|

We therefore have it that:

$$E(S) = \lambda_U \lambda_V (2Tt - t^2)$$

We can now substitute $\lambda_S$ for *E(S)* in our earlier expression for *P(compromise)* to give:

$$P(compromise) = 1 - e^{-\lambda_U \lambda_V (2Tt - t^2)}$$

## Solving for rotation time

We want to know the largest acceptable rotation time $t_{rot}$ for a given risk tolerance *R*. If we replace *t* with $t_{rot}$ and *P(compromise)* with *R*, we can solve for $t_{rot}$:

$$R = 1 - e^{-\lambda_U \lambda_V (2T t_{rot} - t_{rot}^2)}$$

$$\ln(1 - R) = -\lambda_U \lambda_V (2T t_{rot} - t_{rot}^2)$$

$$t_{rot}^2 - 2T t_{rot} - C = 0$$

Where $C = \frac{\ln(1-R)}{\lambda_U \lambda_V}$. Solving using the quadratic formula:

$$t_{rot} = \frac{2T \pm \sqrt{(2T)^2 + 4C}}{2}$$

$$t_{rot} = T \pm \sqrt{T^2 + C}$$

Of these two solutions, one gives us a $t_{rot}$ which is always larger than *T*, which doesn't physically make sense. Noting that $R \in$ ¿, we therefore only consider the solution which always gives a value smaller than *T*:

$$t_{rot} = T - \sqrt{T^2 + \frac{\ln(1-R)}{\lambda_U \lambda_V}}$$

This gives us our maximum acceptable rotation time for a given risk tolerance, for two controls each with a given expected number of vulnerabilities over a given time period.

In order to use this equation to calculate $t_{rot}$ for a given risk tolerance, we need to find representative values for $\lambda_x$ and $\lambda_y$.

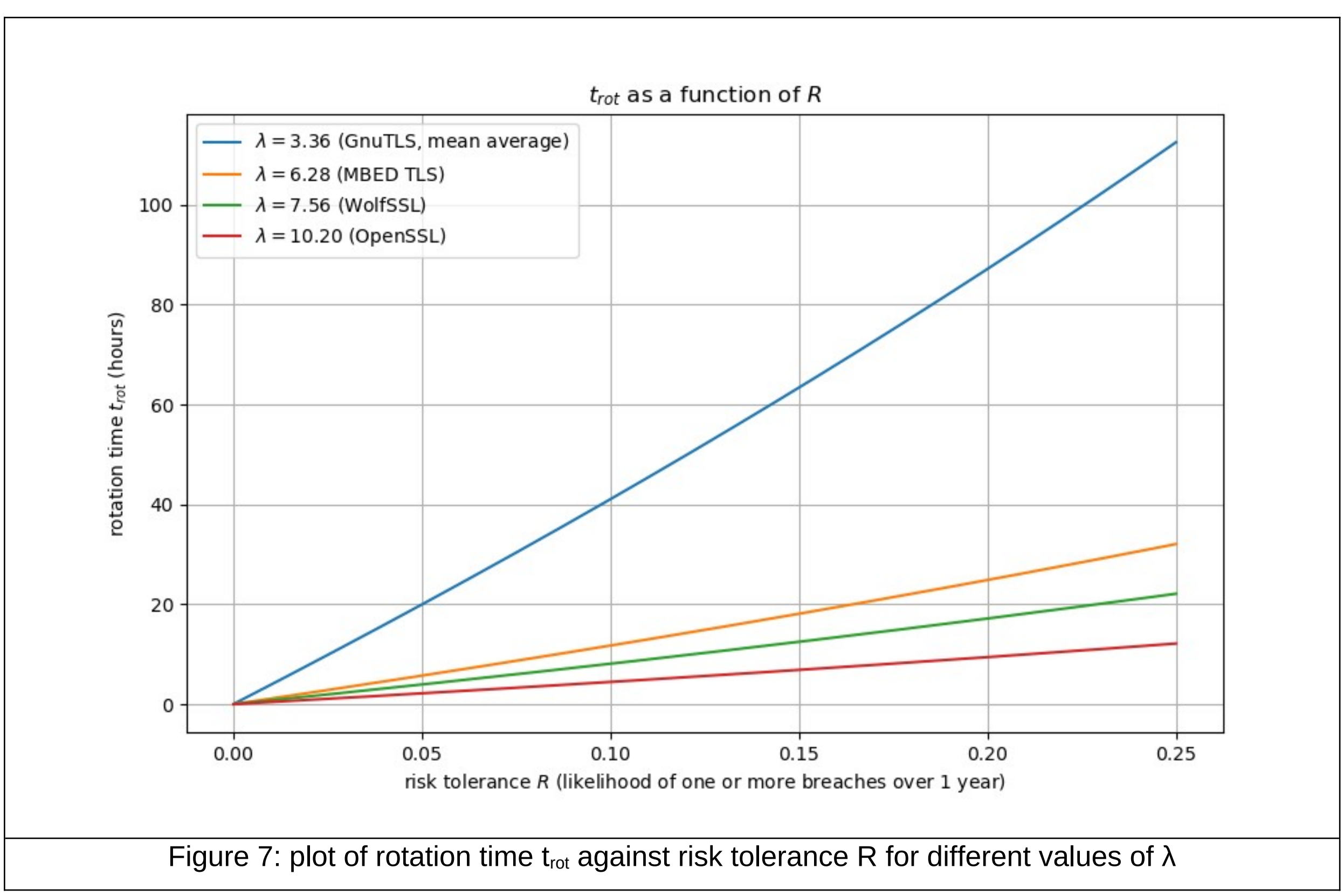

Figure 7: plot of rotation time $t_{rot}$ against risk tolerance R for different values of λ

## Real-world data

To approximate $\lambda_x$ and $\lambda_y$, we sampled vulnerability reports from twelve popular cryptography libraries. Ideally, libraries would be selected based on their level of usage, however accurately measuring usage is challenging. Instead, libraries were selected based on previous sampling sets, which selected libraries with sufficient usage and CVE data [33]. The libraries sampled are listed in Table 1.

| **Cryptography library** | **Primary language** |
|---|---|
| Botan | C++ |
| Bouncy Castle | Java |
| BSAFE | C |

| Cryptlib | C |
|---|---|
| Cryptopp | C++ |
| GnuTLS | C |
| Libgcrypt | C |
| LibreSSL | C |
| MBED TLS | C |
| Nettle | C |
| OpenSSL | C |
| WolfSSL | C |

Table 1: sampled cryptography libraries

Data was retrieved from the National Vulnerability Database (NVD) API maintained by NIST [34].

The full set of reported vulnerabilities present in the NVD was sampled for each library. Each vulnerability has one or more associated CVSS (Common Vulnerability Scoring System) scores [35], with CVSS scores being used to support a quantitative measure of severity [36]. To normalise CVSS scores across all vulnerabilities, the following approach was taken:

- If multiple CVSS versions are used to score a vulnerability: discard all but the newest version.
- If multiple reporting bodies have scored a vulnerability, and the scores differ: discard all but the score provided by NIST.

Plots of annual CVE frequency for each sampled library are given in Appendix A.

Using our earlier assumption that it is reasonable to model vulnerabilities in cryptographic controls as Poisson variables, we approximated $\lambda$ for each library, where $\lambda$ is taken over the time period of 1 year. To be conservative, we calculated approximate values using vulnerabilities of any severity, rather than just including, say, medium or above – though it should be noted that, in practice, the majority of CVEs were medium or higher, so filtering out low-severity CVEs would make little difference to the results. This gave the following spread of values for $\lambda$:

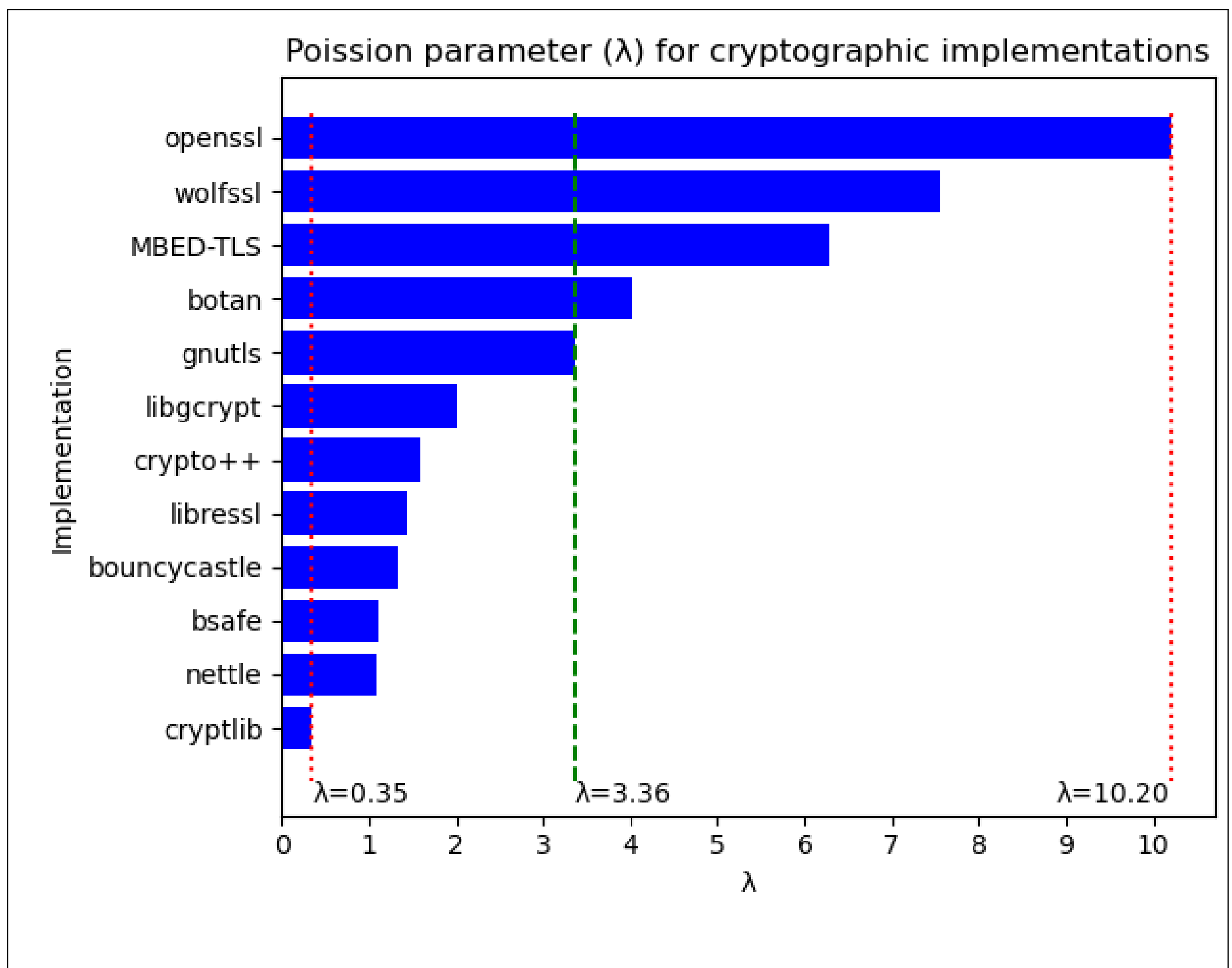

Figure 8: comparison of Poisson parameters λ for different cryptography libraries. The maximum (λ=10.20, openssl), minimum (λ=0.35, cryptlib) and mean average values (λ=3.36) are indicated.

OpenSSL is extremely widely used, and is the subject of rigorous security research, both of which perhaps make its high $\lambda$=10.20 less of a surprise. The mean average across all libraries was $\lambda$=3.36.

Which of these values should we take as representative for $\lambda_x$ and $\lambda_y$? We could use the average across all libraries sampled ($\lambda$=3.36), but in order to remain conservative, it is reasonable to use a value of $\lambda$ for both $\lambda_x$ and $\lambda_y$ that reflects the worst-case scenario, which in this instance would be OpenSSL's $\lambda$=10.20.

The assumption that $\lambda_x=\lambda_y$ also helps simplify our expression for $t_{rot}$:

$$t_{rot}=T-\sqrt{T^2+\frac{\ln(1-R)}{\lambda^2}}$$

A table of values for $t_{rot}$ across different risk tolerances is given as Table B1 in Appendix B.

## Analysis

The representative values in Table B1 indicate that, in order to maintaining the risk posed by emergent vulnerabilities in cryptography components within acceptable tolerances, a rotation interval on the order of hours to days is necessary: the window for responding to emergent vulnerabilities is relatively brief. This is consistent with accepted industry best practices, which recommend that emergency patching follow an accelerated schedule of hours or days [37]. Actual industry response times appear to lag behind this, with a 2020 report [38] finding that more than 50% of organisations are still unable to patch critical vulnerabilities within 72 hours of their release, and around 15% of systems remained unpatched even after 30 days [39]. Many organisations operate universal patching windows for security issues (e.g., “patch within 7 days” [40] [41]). The findings here do not fundamentally challenge established advice; they do, however, highlight the importance of closing the gap between recommendations and practice.

No system can ever be perfectly secure. The methodology in this paper, however, allows architects to design systems that demonstrate an acceptable level of resilience against a quantum-enabled adversary. Since quantum-enabled cryptanalysis has yet to see widespread use in compromising systems, and approaches to designing quantum-resistant systems are still relatively new, it is challenging for architects to exercise “best judgement” in applying design-level controls aimed specifically at mitigating the threat of quantum-enabled cryptanalysis. Our methodology is intended to support that effort, providing approximations for acceptable system tolerances where no proven prior examples might exist.

For systems that operate under demanding risk tolerances, such as national payments infrastructure [42], it is demonstrably useful to approximate technical and operational requirements directly from risk tolerances. The ability to tightly couple system requirements with risk tolerances means that risk mitigations can be designed efficiently, whilst also providing architects with greater confidence that their mitigations are sufficient for purpose. Our approach helps to bridge the gap between abstract risk tolerances and concrete requirements, providing a practical way to approximate the latter from the former.

## Further work

In order to aid analysis, this paper made the conservative simplification that “the system is under constant attack by a motivated adversary; as soon as the attacker is able to exploit a vulnerability in our system, they will do so.” In practice, some vulnerabilities are more likely to be exploited in the wild than others. This simplification might be unnecessary if the analysis were expanded to include the likelihood of each vulnerability being exploited, perhaps using EPSS (exploit prediction scoring system) [43] scores to support where appropriate.

Early analysis performed prior to this paper indicated that using three cryptographic controls in tandem was unlikely to make significant difference to maximum tolerable rotation time. A rigorous treatment that examines the effect of multiple cryptographic controls upon rotation time characteristics would be a useful next step in conclusively describing the relationship between control depth and maximum tolerable rotation time characteristics.

## Conclusion

A security strategy incorporating hybrid encryption and crypto agility is well-positioned to provide robust protection against new vulnerabilities in cryptographic security controls, including those introduced by quantum-capable adversaries. No security measure is completely infallible: while risk can be reduced, it cannot be fully eliminated. Establishing a clearly defined risk tolerance is a vital first step when producing a security architecture, which must then be translated into system requirements. The methodology presented in this paper aims to do precisely that, allowing architects and implementers to derive technical requirements that will keep security risk posed by emergent vulnerabilities within acceptable tolerances, and bridge the gap between technical requirements and organisational policy.

The authors intend for this work to support practitioners and architects seeking pragmatic estimates for system requirements that will ensure their security mitigations remain effective against current and emerging security threats.

## Appendix A: Vulnerability data for cryptography libraries

Annual CVE frequency was plotted for each of the cryptography libraries in Table 1. CVE data was retrieved on the 22nd of October 2025. Two plots are given for each library: one that includes all CVEs, and one that includes only CVEs of medium or higher severity. The plots also coloured to show the frequency of each CVE severity, with the following colour scheme used to denote severity:

| Severity | Colour |
|---|---|
| None/Informational | **Blue** (no vulnerabilities were observed with this severity) |
| Low | **Green** |
| Medium | **Yellow** |
| High | **Orange** |
| Critical | **Red** |

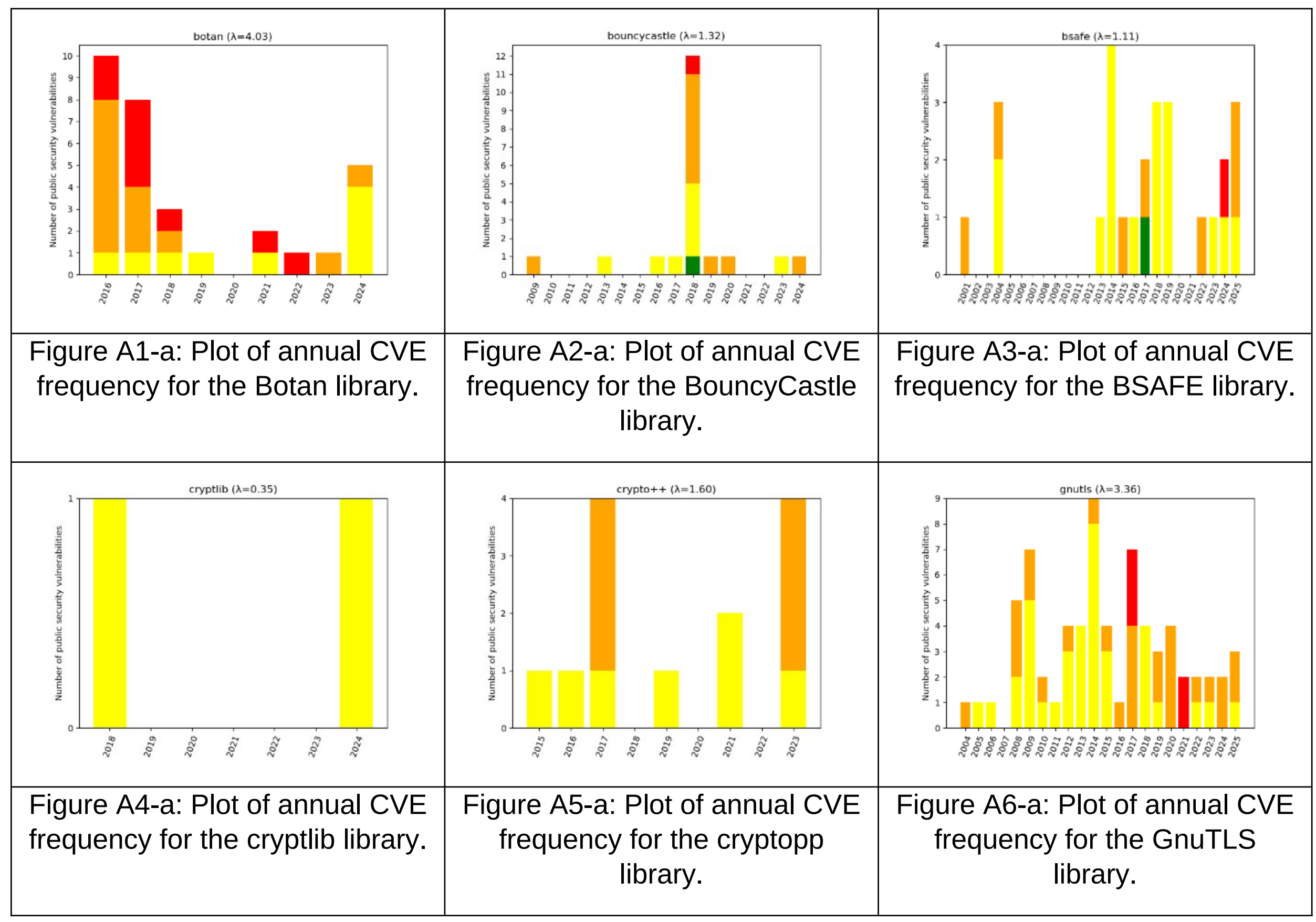

Figure A1-a: Plot of annual CVE frequency for the Botan library.

Figure A2-a: Plot of annual CVE frequency for the BouncyCastle library.

Figure A3-a: Plot of annual CVE frequency for the BSAFE library.

Figure A4-a: Plot of annual CVE frequency for the cryptlib library.

Figure A5-a: Plot of annual CVE frequency for the cryptopp library.

Figure A6-a: Plot of annual CVE frequency for the GnuTLS library.

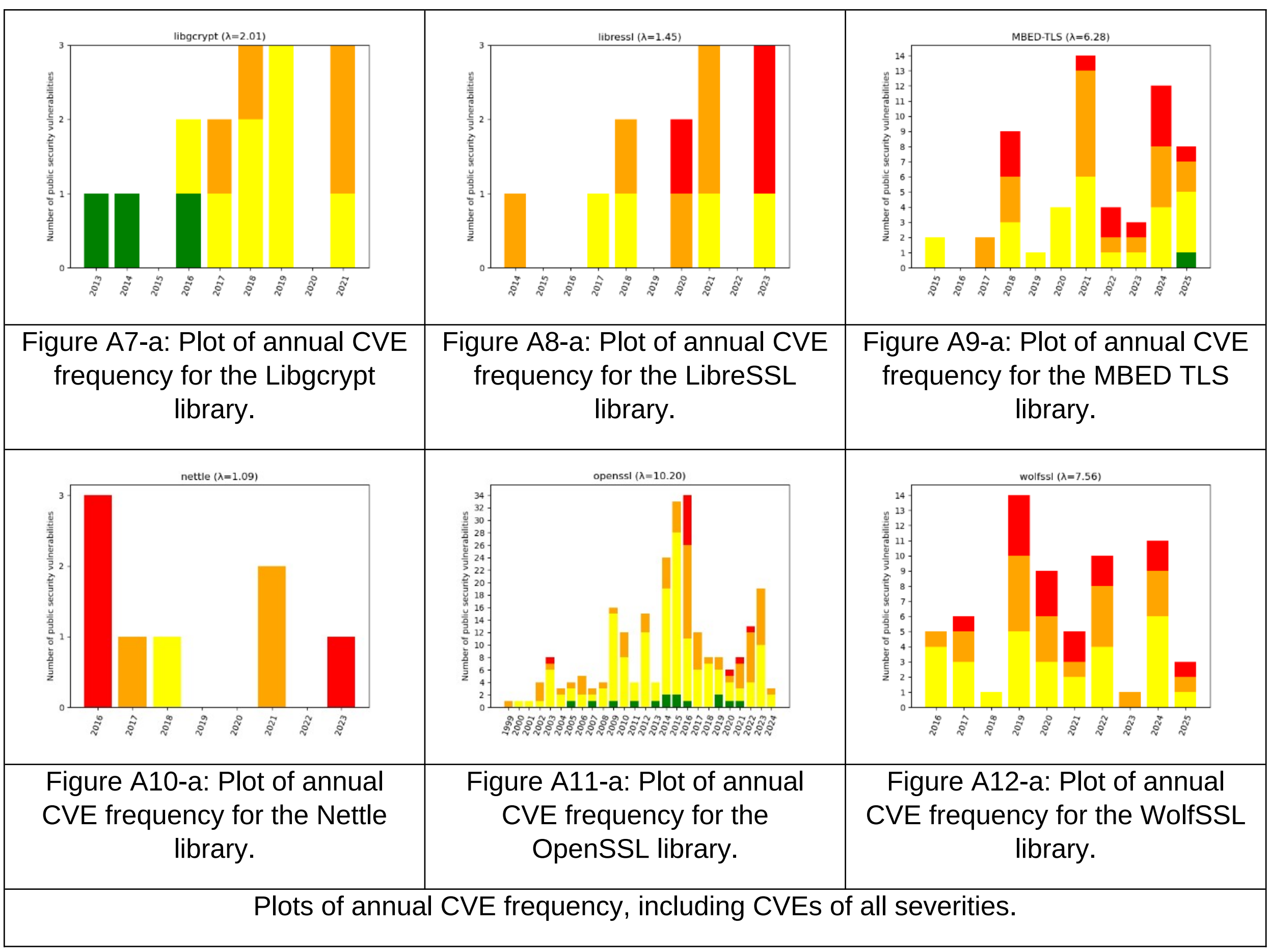

| Figure A7-a: Plot of annual CVE frequency for the Libgcrypt library. | Figure A8-a: Plot of annual CVE frequency for the LibreSSL library. | Figure A9-a: Plot of annual CVE frequency for the MBED TLS library. |
|---|---|---|
| Figure A10-a: Plot of annual CVE frequency for the Nettle library. | Figure A11-a: Plot of annual CVE frequency for the OpenSSL library. | Figure A12-a: Plot of annual CVE frequency for the WolfSSL library. |

Plots of annual CVE frequency, including CVEs of all severities.

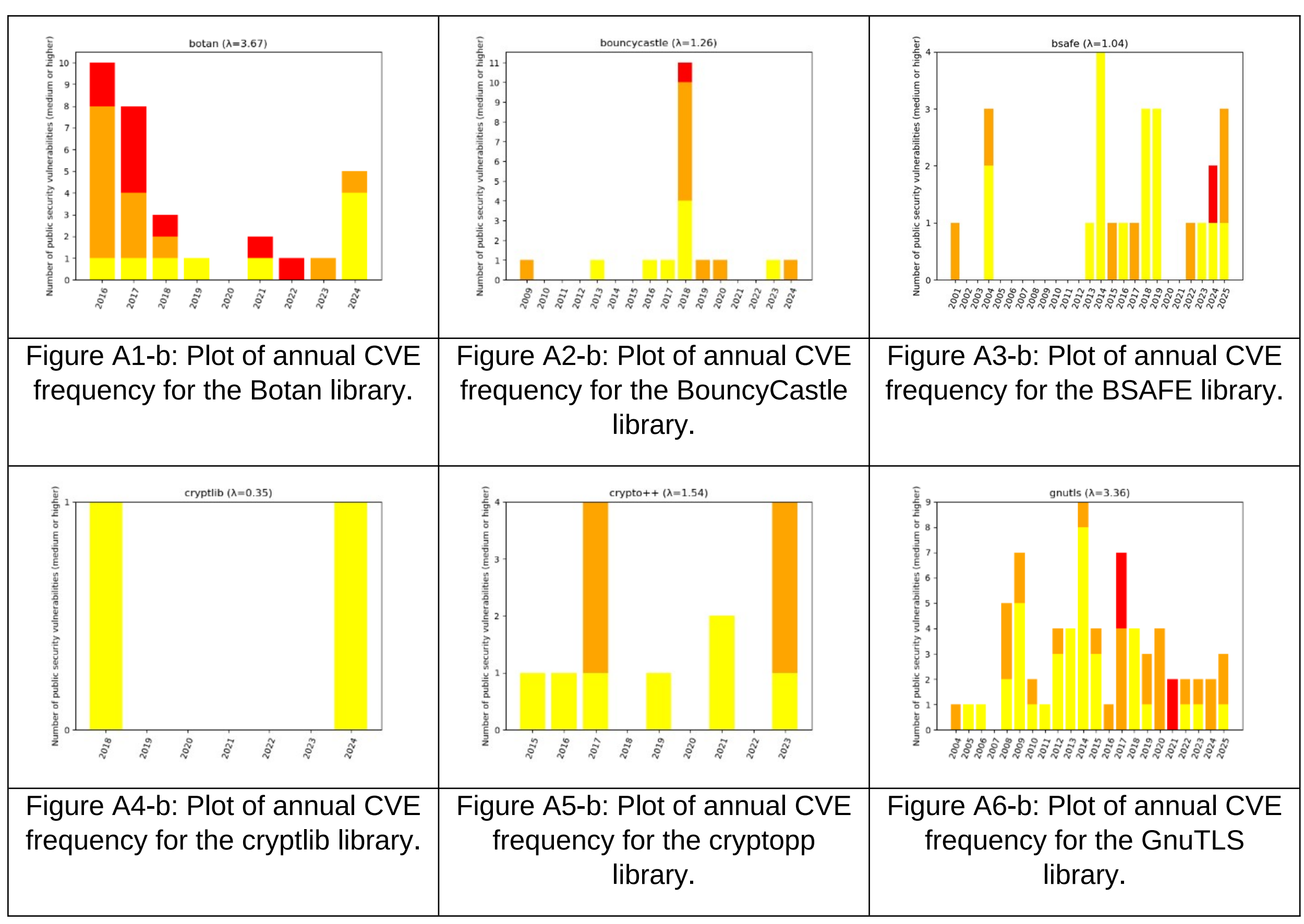

| Figure A1-b: Plot of annual CVE frequency for the Botan library. | Figure A2-b: Plot of annual CVE frequency for the BouncyCastle library. | Figure A3-b: Plot of annual CVE frequency for the BSAFE library. |
|---|---|---|
| Figure A4-b: Plot of annual CVE frequency for the cryptlib library. | Figure A5-b: Plot of annual CVE frequency for the cryptopp library. | Figure A6-b: Plot of annual CVE frequency for the GnuTLS library. |

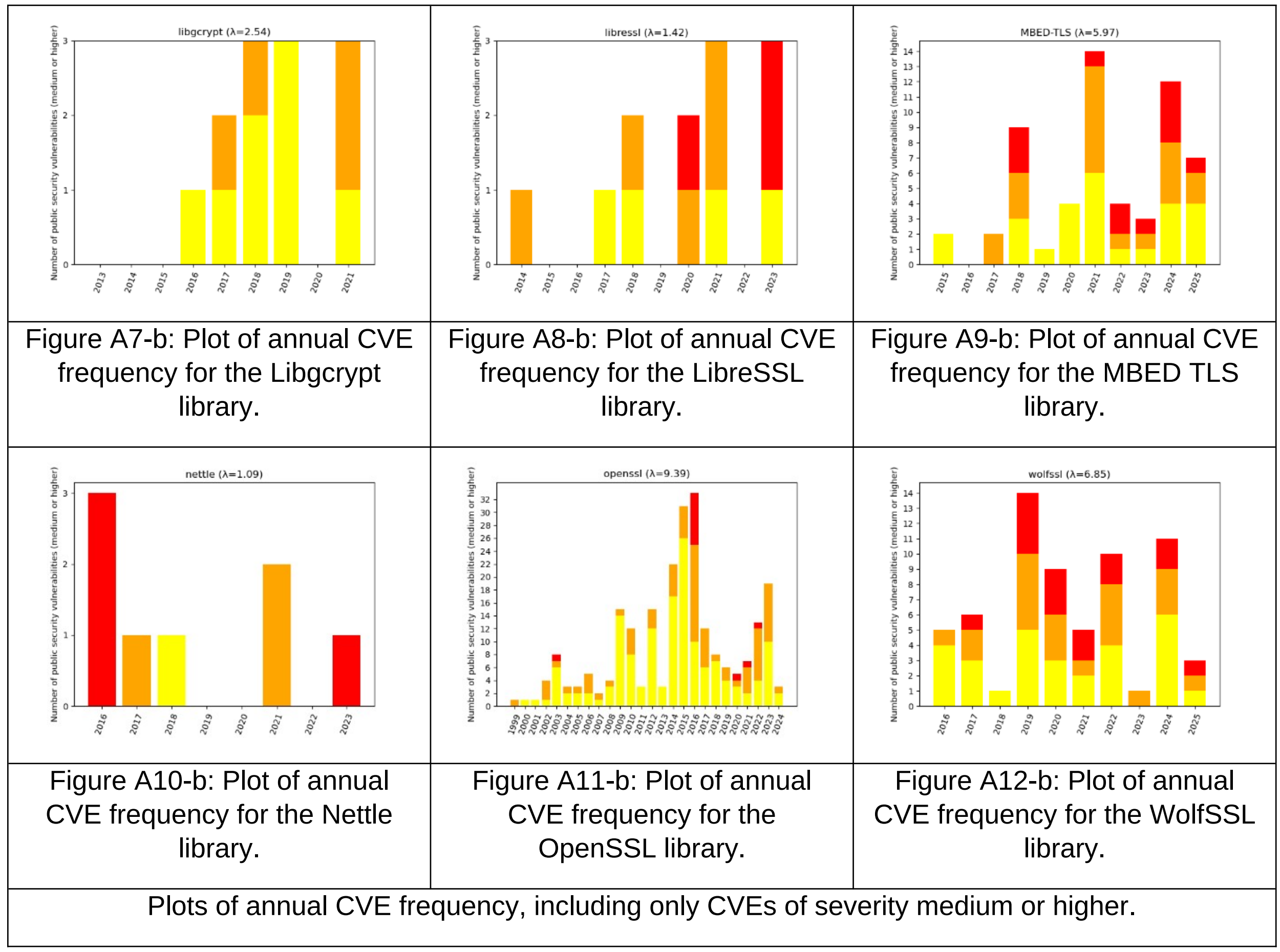

Figure A7-b: Plot of annual CVE frequency for the Libgcrypt library.

Figure A8-b: Plot of annual CVE frequency for the LibreSSL library.

Figure A9-b: Plot of annual CVE frequency for the MBED TLS library.

Figure A10-b: Plot of annual CVE frequency for the Nettle library.

Figure A11-b: Plot of annual CVE frequency for the OpenSSL library.

Figure A12-b: Plot of annual CVE frequency for the WolfSSL library.

Plots of annual CVE frequency, including only CVEs of severity medium or higher.

## Appendix B: Illustrative rotation time

In this paper, we derived an expression for approximating maximum tolerable rotation time:

$$t_{rot} = T - \sqrt{T^2 + \frac{\ln(1-R)}{\lambda_U \lambda_V}}$$

To aid intuition, this expression was used to calculate a set of illustrative values for maximum tolerable rotation over a range of risk tolerances. Risk tolerances range from 1% to 25%, with the intention of covering a range of typical organisational risk tolerances [32]. In the first column, $\lambda_U = \lambda_V = 10.20$, using the Poisson parameter for OpenSSL (the largest across all sampled cryptography libraries), as shown in Figure 8. In the second column, $\lambda_U = \lambda_V = 3.36$, using the mean average Poisson parameter over all sampled cryptography libraries, as shown in Figure 8.

| Risk tolerance (annual likelihood of compromise) | $t_{rot}$ (hours, 2 dp), $\lambda$=10.20 | $t_{rot}$ (hours, 2 dp), $\lambda$=3.36 |
|---|---|---|
| 1% | 0.42 | 3.90 |
| 2% | 0.85 | 7.85 |
| 3% | 1.28 | 11.83 |
| 4% | 1.72 | 15.86 |
| 5% | 2.16 | 19.94 |
| 6% | 2.61 | 24.06 |
| 7% | 3.06 | 28.22 |
| 8% | 3.51 | 32.43 |
| 9% | 3.97 | 36.69 |
| 10% | 4.44 | 41.00 |
| 11% | 4.91 | 45.36 |
| 12% | 5.39 | 49.77 |
| 13% | 5.87 | 54.23 |
| 14% | 6.36 | 58.75 |
| 15% | 6.85 | 63.32 |
| 16% | 7.35 | 67.95 |
| 17% | 7.85 | 72.64 |
| 18% | 8.36 | 77.39 |
| 19% | 8.88 | 82.19 |
| 20% | 9.41 | 87.06 |
| 21% | 9.94 | 92.00 |
| 22% | 10.47 | 97.00 |
| 23% | 11.02 | 102.06 |
| 24% | 11.57 | 107.20 |
| 25% | 12.13 | 112.41 |

Table B1: illustrative values for maximum rotation time, calculated over a range of risk tolerances, using $\lambda_U = \lambda_V = 10.20$ and $\lambda_U = \lambda_V = 3.36$

## Bibliography

[1] D. D. A. D. M. H. N. M. D. M. S. M. a. A. V. Raphael Auer, “Quantum-readiness for the financial system: a roadmap,” *BIS Papers, no. 158,* July 2025.

[2] M. A. A. M. P. Qurban A. Memon, “Quantum Computing: Navigating the Future of Computation, Challenges, and Technological Breakthroughs,” *Quantum Reports,* 2024.

[3] D. G. A. L. S. e. a. Bluvstein, “A fault-tolerant neutral-atom architecture for universal quantum computation.,” *Nature,* 2025.

[4] R. K. S. B. W. H. T. K. G. H. L. J. R. M. T. O. N. C. R. Ryan Babbush, “The Grand Challenge of Quantum Applications,” *arXiv,* 2025.

[5] C. D. H. F. B.-Y. L. L. S. X.-S. T. W. W. G.-M. X. F. Y. H.-F. Y. Y.-S. Z. Y.-R. Z. C.-L. Z. Yao-Yao Jiang, “Advancements in superconducting quantum computing,” *National Science Review,* vol. 12, no. 8, 2025.

[6] D. J. &. L. T. Bernstein, “Post-quantum cryptography : dealing with the fallout of physics success,” *Cryptology ePrint Archive,* p. 20, 2017.

[7] A. S. L. A. R. L. Rivest, “A Method for Obtaining Digital Signatures and Public-Key Cryptosystems,” *Communications of the ACM,* 1978.

[8] V. R. Joan Daemen, “AES Proposal: Rijndael,” 1999.

[9] P. W. Shor, “Polynomial-Time Algorithms for Prime Factorization and Discrete Logarithms on a Quantum Computer,” *Proceedings of the 35th Annual Symposium on Foundations of Computer Science,* 1994.

[10] L. K. Grover, “A Fast Quantum Mechanical Algorithm for Database Search,” in *Proceedings of the 28th Annual ACM Symposium on Theory of Computing*, 1996.

[11] R. A. Grimes, Cryptography Apocalypse: Preparing for the Day When Quantum Computing Breaks Today's Crypto, Wiley, 2019.

[12] Deloitte Center for Integrated Research, “Quantum computing over the next five years: Scenario planning for strategic resilience,” 2025.

[13] A. R.-P. a. N. C. a. T. Finogina, “An electoral exception? Quantum computing-readiness and internet voting,” *eJournal of eDemocracy and Open Government,*

vol. 16, no. 3, 2024.

[14] G. e. a. Alagic, “Status report on the third round of the NIST post-quantum cryptography standardization process.,” *NIST,* p. 90, 2022.

[15] National Institute of Standards and Technology, “Announcing Issuance of Federal Information Processing Standards (FIPS) FIPS 203, Module-Lattice-Based Key-Encapsulation Mechanism Standard,” Federal Register, Washington, D.C., 2024.

[16] A. R. G. C. L. D. e. a. Cano Aguilera, “ In-line rate encrypted links using pre-shared post-quantum keys and DPUs,” *Scientific Reports,* 2024.

[17] M. G. M. M. a. T. S. S. Jose Deodoro, “Quantum Computing and the Financial System: Spooky Action at a Distance?,” *IMF Working Paper,* no. WP/21/71.

[18] National Cyber Security Centre, “Annual Report 2025,” p. 27.

[19] D. Bhasker, “Decrypting the Future: Insights from RSAC2025 Cryptographers’ Panel,” *ISC2 Insights,* 14 May 2025.

[20] N. a. B. J. a. F. M. a. G. B. a. S. D. Bindel, “Hybrid Key Encapsulation Mechanisms and Authenticated Key Exchange,” in *Post-Quantum Cryptography*, Springer International Publishing, 2019, pp. 206-226.

[21] European Telecommunications Standards Institute, “Quantum Safe Cryptography and Security,” 2015.

[22] C. Naether, “Toward a Common Understanding of Cryptographic Agility – A Systematic Review,” IEEE Dataport, 2025.

[23] C. L. C. D. M. D. R. A. S. M. N. B. H. R. T. S. B. W. K. Barker E, “Considerations for Achieving Crypto Agility: Strategies and Practices,” National Institute of Standards and Technology, 2025.

[24] N. A. a. N. S. a. A. W. a. A. H. a. T. Grasmeyer, “On the State of Crypto-Agility,” Cryptology {ePrint} Archive, Paper 2023/487, 2023.

[25] A. A. Fall, “SoK: Systematizing Hybrid Strategies for the Transition to Post-Quantum Cryptography,” *Cryptology ePrint Archive,* 2025.

[26] G. C. a. S. S. a. P. H. a. S. B. a. S. Das, “Post-Quantum Cryptography and Quantum-Safe Security: A Comprehensive Survey,” *arXiv preprint*

*arXiv:2510.10436,* 2025.

[27] R. Anderson, “Security in open versus closed systems—the dance of Boltzmann, Coase and Moore,” Cambridge University.

[28] E. Rescorla, “Is finding security holes a good idea?,” *IEEE Security & Privacy,* vol. 3, no. 1, pp. 14-19, 2005.

[29] Y. K. M. Omar Alhazmi, “Prediction capabilities of vulnerability discovery models,” in *Annual Reliability and Maintainability Symposium*, 2006.

[30] K. O. John D Musa, “A logarithmic Poisson execution time model for software reliability measurement,” in *Proceedings of the 7th international conference on Software engineering*.

[31] R. J. Anderson, “26.2.4 Assurance Growth,” in *Security engineering: a guide to building dependable distributed systems*, John Wiley & Sons, 2010.

[32] National Institute of Standards and Technology, “Identifying and Estimating Cybersecurity Risk for Enterprise Risk Management,” Federal Register, Washington, DC, 2021.

[33] J. Blessing, M. A. Specter and D. J. Weitzner, “Cryptography in the Wild: An Empirical Analysis of Vulnerabilities in Cryptographic Libraries,” MIT, 2024.

[34] National Institute of Standards and Technology, 22 October 2025. [Online]. Available: https://nvd.nist.gov/.

[35] National Institute of Standards and Technology, “Vulnerability Metrics,” 22 October 2025. [Online]. Available: https://nvd.nist.gov/vuln-metrics/cvss.

[36] FIRST (Forum of Incident Response and Security Teams), “Common Vulnerability Scoring System v4.0 Specification Document,” Forum of Incident Response and Security Teams, 2023.

[37] K. S. Murugiah Souppaya, “Guide to enterprise patch management planning,” *Special Publication (NIST SP),* 2022.

[38] Automox, “2020 cyber hygiene report: What you need to know now - lessons learned from a survey of the state of endpoint patching and hardening,” Automox, 2020.

[39] A. J. M. Z. M. A. B. Nesara Dissanayake, “Software Security Patch Management -- A Systematic Literature Review of Challenges, Approaches, Tools and Practices,” 2020.

[40] S. e. a. Narang, “Bi-criterion problem to determine optimal vulnerability discovery and patching time.,” *International Journal of Reliability, Quality and Safety Engineering,* 2018.

[41] Y. Roumani, “Patching zero-day vulnerabilities: an empirical analysis,” *Journal of Cybersecurity,* vol. 7, no. 1, 2021.

[42] Bank of England, “The digital pound: Technology Working Paper,” Bank of England, 2023.

[43] J. J. a. S. R. a. B. E. a. M. R. a. I. Adjerid, “Exploit Prediction Scoring System (EPSS),” *CoRR,* vol. abs/1908.04856, 2019.